\documentclass[16pt]{article}
\usepackage[pdftex]{graphicx}
\title{Pricing American and Asian Options}
\author{Pat Muldowney}

\newtheorem{example}{Example}

\newcommand{\vt}{\vspace{5pt}\\}
\newcommand{\R}{\mathbf{R}}

\newcommand{\E}{{\mathrm{{E}}}}

\newcommand{\N}{\mathcal{N}}
\newcommand{\I}{\mathbf{I}}

\newcommand{\sq}{\simeq}

\newcommand{\bN}{\mathbf{N}}

\newcommand{\pr}{\mathbf{P}}

\newtheorem{maple}{Calculation}

\usepackage[pdftex]{graphicx}

\begin{document}
\date{}\maketitle

\begin{abstract}
An analytic method for pricing American call options is provided; followed by an empirical method for pricing Asian call options. The methodology is the pricing theory of \cite{MTRV}, which is also the source of the notation.
\end{abstract}

Suppose a share has market value $z(s)$ at time $s$, and suppose an option contract, between two parties, is entered into at time $s=0$, so that one of the parties to the contract acquires the right, but not the obligation, to purchase the share, from the other party, for a fixed price $\kappa$ at any time $t$, $0<t\leq \tau$. The option expires at time $\tau$, so there is no right to purchase the share at times $t > \tau$. 
These are the features of an American call option.

An {American call option} can be exercised at any time during the term or lifetime of the contract. This differs from a European call option which can only be exercised on the date of termination of the contract.

Here are some questions about American call options. What is the initial value (or purchase price) $w(0)$ of the option contract at time $0$? At any time $t$, $0<t \leq \tau$, what is the market value $w(t)$ of the (unexercised) call option?

If time $t>\tau$, then $w(t)=0$ since the option can no longer be exercised. At time $t=\tau$, the American call (if not previously exercised) has value
\[
w(\tau) = \max\{z(\tau) - \kappa,\,0\},
\]
the same as the corresponding European call option.

At time $ t<\tau$, the American call option and the corresponding European call option have non-zero value\footnote{If $\kappa$ is excessively large the option may be effectively worthless. See Figure \ref{fig:figure4} below.} even if $z(t) < \kappa$. The difference between the two is that, for $t<\tau$, the European call  cannot be exercised even if $z(t)>\kappa$.

If $z(0)$ is the observed value of the underlying share at the time of writing the contract, the purchase price $w(0)$ and the exercise price $\kappa$ of this contract must satisfy
\begin{equation}\label{Exercise Immediately!}
w(0)+ \kappa  \geq z(0).
\end{equation}
The reason for this is that, if $w(0) < z(0) - \kappa$, the option can be exercised immediately, the share can be purchased for the amount $\kappa$, and then re-sold for amount $z(0)$, realizing a riskless profit of $z(0) -\kappa - w(0)$.

For $t<\tau$  the American call option will not be exercised at time $t$  if $z(t)<\kappa$. What if $z(t)>\kappa$? In that case, and disregarding the  outlay required to purchase the option, the holder of the option will obtain immediate profit of amount $z(t)-\kappa$ by purchasing the share for the option exercise price $\kappa$ and immediately re-selling the share for the market price $z(t)$. If $t=\tau$ that is what is expected to happen, since there are no further opportunities to purchase the share at option exercise price $\kappa$.

But if $t<\tau$, and even if $z(t)>\kappa$, then, depending on various factors, the holder of the American call may decide \textbf{not} to exercise in case they expect the underlying share price to rise further in the time interval $]t, \tau]$. If that were to happen, then the profit (or \emph{payoff}) $z(s)-\kappa$ to be obtained by exercising the option at the later date $s>t$ would be greater than what would be obtained by exercising at time $t$. 

This introduces a new element of unpredictability. If time $t$ is the ``present'', the share price $z(t)$ has a known or observed value. Call this value $\kappa_t$. Even if  $\kappa_t>\kappa$, 
the option holder, in anticipation of further increase in the price of the underlying share, may decline to exercise the option at time $t$.   

On the face of it, in advance of exercising, the optimal time $t$ at which to consider exercising the American call is the time $\tilde{t}$ for which the share price $z(t)$ is maximal, $0<t\leq \tau$; so $x(\tilde t) > x(s)$ for $s \neq \tilde t$. 

However, the time value of money has to be factored into this argument. The time value of money is measured by the risk-free interest rate $\rho$. The following formulation takes account of time value of any potential option payoff.

If $\tau \geq \tilde t > s$ and $z(s)> z(\tilde t) > \kappa$, it is more advantageous to exercise the American call at the earlier time $s$ if
\[
\left(z( s) - \kappa \right) e^{\rho(\tilde t - s)} >
z(\tilde t) - \kappa.
\]
At time $s=0$ the optimal\footnote{Likewise, at times $s>0$ the optimal exercise time $\varsigma$ is unknown, $s\leq \varsigma \leq \tau$.}  exercise time $\varsigma$ is unpredictable.
It is possible that the holder of an American call will exercise the option in advance of the optimal time $\varsigma$,  or will do so after the optimal point has passed. Or will not exercise at all.

 Assuming ``equal likelihood'', each of these possibilities is just as likely as the other. On this reasoning, the exercise time (or date) $ t$ of an American call option is a random variable whose \emph{expected value} is $\varsigma$.
 

To explore further, assume that $t<\tau$ is the ``present'' so the share price $z(t) = \kappa_t$ is known. Suppose that, even if $\kappa_t>\kappa$,  the option holder, in anticipation of further increase in the price of the underlying share, declines to exercise the option at time $t$. Assume that the likelihood of the option holder making this decision   is equal to the probability that $z(s)>\kappa_t$ for some $s \in \,]t,\tau]$. That is,
\[
z(s) - \kappa > z(t) - \kappa = \kappa_t - \kappa.
\]
In reality it is necessary to factor in the time value of money here. The option holder will decline to exercise if they expect that the discounted value of the term on the left hand side of the following inequality exceeds the right hand side:
\[
\left(z(s) - \kappa\right) e^{-\rho (t-s)} > z(t) - \kappa = \kappa_t - \kappa.
\]
If $t$ is given, and if $s$ is the ``present'', with $s<t$, similar considerations apply to any decision  to exercise the American call during the period $]s,t]$.

This argument implies certain assumptions about the motivation and behaviour of option holders/buyers as a body, in regard to future prices of an underlying share. But the future market value of the underlying share is said to be ``correctly'' established by similar motivation and behaviour of those buying and selling the  share. So there seems to be some justification for such assumptions.

Take time $0$ to be the ``present''. Suppose $s$ is a given ``future'' time, 
$0<s <\tau$; so it is not known whether $z(s)$ is going to be less than or greater than the exercise price $\kappa$. 
Consider  an option exercise decision at the future time $s$. 

For any exercise date $s$ ($0<s \leq \tau$) denote the discounted ``difference'' function for the option by
\begin{equation}\label{discounted difference function 1}
p(s,z(s)) := \left(z(s) - \kappa\right) e^{-\rho s}.
\end{equation}
If time $t>0$ is the present and $0<t<s \leq \tau$, then the relevant difference function is discounted from time $s$ to time $t$:
\begin{equation}\label{discounted difference function 2}
p_t(s,z(s)) := \left(z(s) - \kappa\right) e^{-\rho (s-t)}.
\end{equation}
Take time $0$ to be the present, so option exercise decisions are based on (\ref{discounted difference function 1}). The option is exercisable (or ``in the money'') if $z(s) > \kappa$ for some $s \in\,]0,\tau]$; and it is optimally exercisable at time $s=s({z_T})$ if, for some other time $s'$ before or after $s$, the payoff, discounted from time $s'$ to time $0$, is not greater than the payoff at time $s({z_T})$ discounted to time $0$.
The optimal exercise time $s({z_T})$, if exercise\footnote{Exercise may never be feasible if the exercise price $\kappa$ is set too high. See Figure 4 below.} is feasible, is variable for different histories $z_T$.

The graph or history $z_T$ is a potential occurrence of a joint-basic
 observable $Z_T$ if there is a distribution function $F$ for 
which\footnote{It is sometimes assumed, incorrectly, that $F$ is the geometric Brownian distribution function $\mathcal G$.}
\[
Z_T \sq z_T\left[\R_+^T,F\right].
\]
In that case, for $0< s\leq \tau$ the discounted payoff function values $p( s,z_T)$ are potential occurrences of a contingent observable
\[
p( s,Z_T)\sq p( s, z_T) \left[\R_+^T,F\right],
\]
whose expected value is 
\begin{equation}
\label{expected discounted payoff}
q( s) := \E\left[p\left( s,Z_T\right)\right] 
= \int_{\R_+^T} p( s,z_T)F(I[N]).
\end{equation}
If the exercise date $ s$ is selected, then $ s$ is fixed in this integration. Therefore the point-integrand $p$ is a cylinder function, so the integral reduces to a one-dimensional integral\footnote{If $F$ is the geometric  Brownian distribution function $\mathcal G^{\mu\sigma}$ then the integral is the one given in line 5 of \cite{MTRV} page 442, but with natural process growth rate $\mu$ in place of the risk-free growth rate $\rho$, and with $p(s,z_T)$ in place of $\max\{z_\tau - \kappa,0\}$.} on $\R_+ = \R_+^{\{ s\}}$.

Choose $ s$ so that $q( s)$ is maximized\footnote{Set aside, for the moment, the question of how to make this choice.} and denote the chosen $ s$ by $\varsigma$. Then $0 \leq\varsigma \leq \tau$, and, if time $0$ is the present, then, at time $0$, it is predicted that the American call option will be exercised at time $\varsigma$.

The question also arises as to  whether, if  the ``present'' is some time $t$ later than $0$, the predicted exercise date $\varsigma_t$ will  be different.

The object of this article is to consider ways of finding a purchase price $w(t), = w_t,$ for a call option at any time $t$, $0 \leq t < \tau$; in particular the price $w(0), = w_0$, payable by the initial purchaser of the option. Consider the latter problem first, so $t=0$ is the present.

For a European call option the exercise date is $\tau$, and the initial value of the option is obtained by means of the risk-neutral pricing formula (8.46) of \cite{MTRV}, page 440:
\[
e^{-\rho \tau} \int_{\R_+^T} f_\tau(z_T) \bar F(I[N]),
\]
where $\bar F$ is the risk-neutral martingale distribution function discussed in Sections 8.14 and 8.15, pages 436--440 of \cite{MTRV}, and $f_\tau(z_T) = \max\{z(\tau)-\kappa, 0\}$.

For the corresponding American call option, the predicted exercise date is $\varsigma$, which may be less than $\tau$. But apart from that, the pricing argument is the same and gives initial value of the American call as 
\[
e^{-\rho \varsigma} \int_{\R_+^T} f_{\varsigma}(z_T) \bar F(I[N]),
\]
where  $f_{\varsigma}(z_T) = \max\{z(\varsigma), 0\}$. The integral on $\R_+^T$ can be replaced by one on $\R_+^{]0, \varsigma]}$. Either way, since the integrand is cylindrical the integral reduces to a one-dimensional integral on $\R_+, =\R_+^{\{\varsigma\}}$.


When it is assumed\footnote{This assumption is generally incorrect. See \cite{MTRV} Section 9.9, pages 479--485.} that the underlying asset price process $Z_T$ is geometric Brownian with $F=\mathcal G^{\mu \sigma}$, then Section 8.16 (\cite{MTRV} pages 440--444) shows that $\bar F=\mathcal G^{\rho \sigma}$, where $\sigma$ is the volatility, $\mu$ is the growth rate of the process, and $\rho$ is the risk-free interest rate. In that case the initial value of the American call is given by (8.53) of \cite{MTRV} page 442, with $\tau$ replaced by $\varsigma$.

How can the optimal exercise time $\varsigma$ be established? 
If the process distribution function $F$ is an  expression such as $\mathcal G^{\mu\sigma}$ then the 
expected discounted difference $q( s)$ of (\ref{expected discounted payoff}) has a particular form whose maximum value can be estimated; so the corresponding $ s$, $= \varsigma$, can also be estimated.

On the other hand, suppose $F$ (or, at least, approximate values of $F$) is determined empirically, as indicated in \cite{MTRV} Section 9.11 page 488. Then it may be possible  to calculate empirically the {expected discounted difference} $q( s)$ of (\ref{expected discounted payoff}) using different values of $ s$, and use the results to estimate $\varsigma$, the exercise date which gives the largest {expected discounted difference}.

As an exercise in determining optimal exercise time $\varsigma$, take $F=\mathcal G^{\mu\sigma}$, which is the traditional (but generally incorrect) assumption. Then, 
using Lemma 33 (pages 309--310 of \cite{MTRV}) with change of variable $z( s) = e^u$, (\ref{expected discounted payoff}) becomes
\begin{equation}\label{expected difference 1}
q(s) = \int_{\R_+^{]0, s]}} e^{- s}\left(z( s) - \kappa\right) \mathcal G^{\mu\sigma}(I[N]);
\end{equation}
so 
\begin{eqnarray*}
q(s)&=& e^{- s}\left(\int_{-\infty}^\infty  \frac {e^u}{\sigma \sqrt{2\pi  s}} e^{-\frac 12 \left(\frac{u-\mu}{\sigma \sqrt {s}}\right)^2 } du
-\kappa \int_{-\infty}^\infty  \frac 1{\sigma \sqrt{2\pi s}} e^{-\frac 12 \left(\frac{u-\mu}{\sigma \sqrt {s}}\right)^2 } du\right)\vt
&=& e^{-s} \left(e^{\mu s + \frac{\sigma^2 s}2} - \kappa\right).
\end{eqnarray*}
With $\mu=0$, $\sigma=0.5$, $\tau=2$, and with exercise price $\kappa=0.1$, $2$, $1$, and $5$, Figures 1, 2, 3, and 4, respectively illustrate the expected values $q(s)$ of the corresponding options if they are exercised at times $s$ for $0\leq s \leq 2$.

\begin{figure}
\begin{minipage}[b]{0.45\linewidth}
\centering
\includegraphics[width=\textwidth]{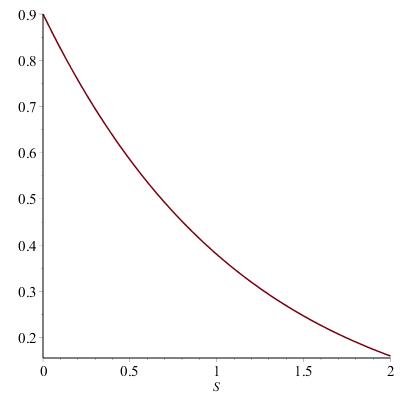}
\caption{$\kappa=0.1$}
\label{fig:figure1}
\hspace{0.5cm}
\centering
\includegraphics[width=\textwidth]{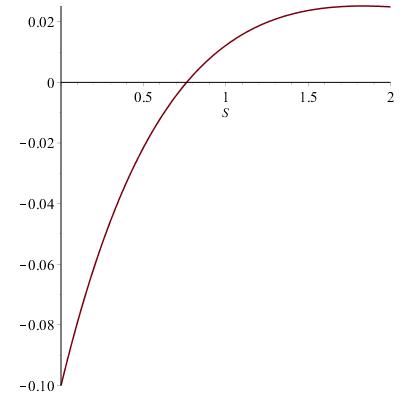}
\caption{$\kappa=1.1$}
\label{fig:figure2}
\hspace{0.5cm}
\centering
\includegraphics[width=\textwidth]{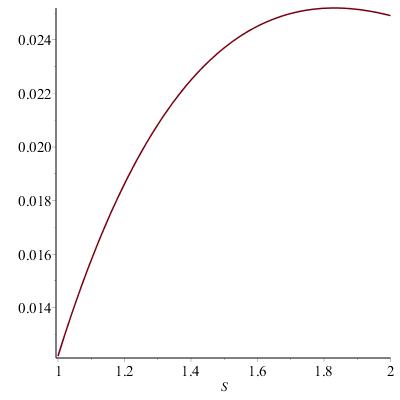}
\caption{$\kappa=1.1$, $t=0$, $\tau=2$}
\label{fig:figure2++}

\end{minipage}
\hspace{0.5cm}
\begin{minipage}[b]{0.45\linewidth}
\centering
\includegraphics[width=\textwidth]{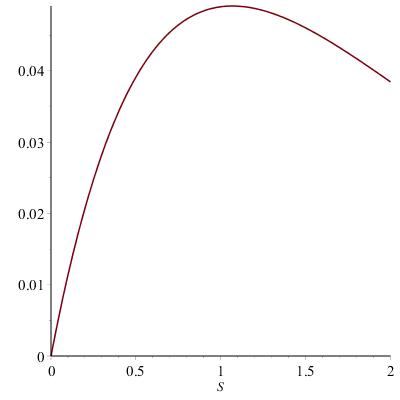}
\caption{$\kappa=1$}
\label{fig:figure3}
\hspace{0.5cm}
\centering
\includegraphics[width=\textwidth]{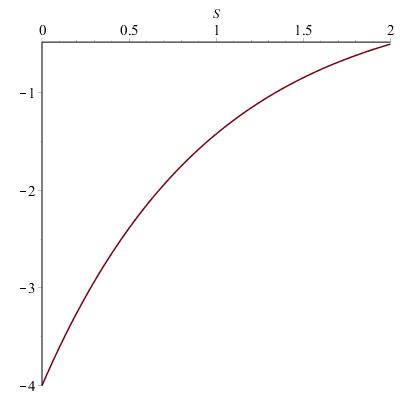}
\caption{$\kappa=5$}
\label{fig:figure4}
\hspace{0.5cm}
\centering
\includegraphics[width=\textwidth]{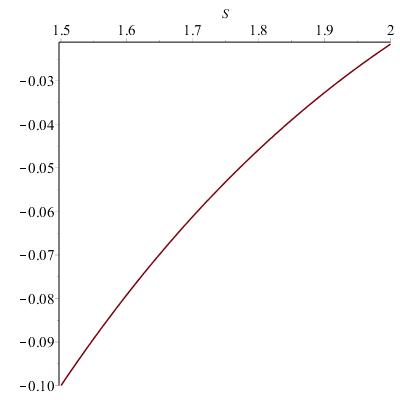}
\caption{$\kappa=1.1$, $t=1.5$, $\tau=2$}
\label{fig:figure5}
\end{minipage}
\end{figure}

These graphs indicate that if exercise price $\kappa$ is very low  (Figure 1), the situation can be compared to that of (\ref{Exercise Immediately!}), and it is best to exercise the option immediately. 
In other words, at time $t=0$ essentially riskless profit is available ``here and now''; so why  wait  around for potential increases in the underlying share price? After all, the latter could actually decrease.

If $\kappa$ is a bit larger (Figure 2), at time $t=0$ it is safest to plan on exercising the option at or 
near\footnote{Figure \ref{fig:figure2++} gives a more detailed graph of $q(s)$ for this case. It shows that the maximal value of $q(s)$ occurs at about $s=1.83$.}  the contract termination date $s=\tau$. Because, even if (as in the cases portrayed here) the growth rate $\mu$ of the underlying share price is zero, there is a ``structural'' mean growth rate of $\frac 12 \sigma^2$ due to volatility. So the attitude of the investor in American call options at time $t=0$ will lean\footnote{Some commentary on American call options suggests that there is no merit at all in early exercise.} towards late exercise.

For intermediate values of $\kappa$ (Figure 3) the optimal exercise time $\varsigma \approx 1$ is intermediate between $0$ and $2$. And for very large $\kappa$ (Figure 4), the long position in the option contract appears unattractive---potential buyers may not be too interested in the long position in an option with such a high exercise price, and will not pay very much for it; so it can only be sold for some very small amount of money. 

Figure 4 does not imply that an option with these parameters can never be in the money. The function $q(s)$ yields expected rather than actual values. But the diagram suggests that the option should be immediately exercised if and when it achieves in-the-money status, without waiting around for further increases in underlying share price. Secondly, at time $t=0$ the purchase price of such an option should be based on exercise at time $s=\tau$, not earlier, in order to allow for any possible growth in underlying share price.

The optimal exercise date $\varsigma$ 
can be estimated by methods such as the one demonstrated in Figures 1 to 4. Once $\varsigma$
is established, the Black-Scholes pricing formulae (8.53)  of \cite{MTRV} page 442 can be used, with $\varsigma$ replacing $\tau$; giving the initial ($t=0$) value of the American call option as
\begin{equation}\label{w0}
w_0= z_0 \bN^{0,1}
\left( \frac{\ln \frac {z_0}{\kappa} + \left( \rho + \frac 12 \sigma^2\right) \varsigma}{\sigma \sqrt{\varsigma}} \right)
-\kappa e^{-\rho \varsigma} \bN^{0,1}
\left( \frac{\ln \frac {z_0}{\kappa} + \left( \rho - \frac 12 \sigma^2\right) \varsigma}{\sigma \sqrt{\varsigma}} \right)
\end{equation}
Now suppose the option, created at time $s=0$, is unexercised at time $t$, $0<t <\tau$. Suppose the holder of the option wishes to sell it at time $t$. What price should be charged for it? In other words, what is the value $w(t)$ (or $w_t$)?

Can it be expected that, at time $t$ later than $0$, the expected exercise date for the as yet unexercised option is $\varsigma$, the exercise date obtained by maximising $q(s)$ in (\ref{discounted difference function 1})? If that were the case, then formula (8.54) of \cite{MTRV} page 442 could be used for $w_t$, with $\varsigma$ replacing $\tau$; so $w_t$ would be
\begin{equation}\label{(8.54)}
  z_t \bN^{0,1}
\left( \frac{\ln \frac {z_t}{\kappa} + \left( \rho + \frac 12 \sigma^2\right) \left(\varsigma -t \right)}{\sigma \sqrt{\varsigma-t}} \right)
-\kappa e^{-\rho \left(\varsigma -t \right)} \bN^{0,1}
\left( \frac{\ln \frac {z_t}{\kappa} + \left( \rho - \frac 12 \sigma^2\right) \left(\varsigma -t \right)}{\sigma \sqrt{\varsigma-t}} \right)
\end{equation}
The trouble with this argument is that $\varsigma$ is not necessarily the actual exercise date. It is a date estimated
at time $s=0$ for the purpose of pricing the option at that time. The option holder may choose to exercise at some time $s$ earlier or later than $\varsigma$. 

Indeed, if exercise is delayed, then $t$ (which is the ``present'' in the new pricing scenario) may be greater than $\varsigma$ in (\ref{(8.54)}).  In that case (\ref{(8.54)}) makes no sense. 

However, with appropriate modification the pricing argument may be adapted for $t >0$. At time $t\geq 0$ the estimated exercise date $\varsigma_t$ is the value of $s$ ($t \leq s \leq \tau$)  which maximises
\begin{equation}
\label{expected discounted payoff 2}
q_t( s) := \E\left[p_t\left( s,Z_T\right)\right] 
= \int_{\R_+^{]t, s]}} p_t( s,z_T)F(I[N]).
\end{equation}
If the distribution function $F$ for the share price process $Z$ is $\mathcal{G}^{\mu\sigma}$, (\ref{expected discounted payoff 2}) becomes
\begin{equation}\label{expected difference 2}
q_t(s) = \int_{\R_+^{]t, s]}} e^{- (s-t)}\left(z( s) - \kappa\right) \mathcal G^{\mu\sigma}(I[N]).
\end{equation}
As before, since the point-integrand $p_t$ is cylindrical, this gives
$q_t(s) =$
\[
e^{- (s-t)}\left(\int_{-\infty}^\infty  \frac {e^u}{\sigma \sqrt{2\pi  (s-t)}} e^{-\frac 12 \left(\frac{u-\mu}{\sigma \sqrt {s-t}}\right)^2 } du
-\kappa \int_{-\infty}^\infty  \frac 1{\sigma \sqrt{2\pi (s-t)}} e^{-\frac 12 \left(\frac{u-\mu}{\sigma \sqrt {s-t}}\right)^2 } du\right)
\]
\[
\mbox{or }\;\;\;\;\;\;\;\;\;\;\;q_t(s) =
e^{-(s-t)} \left(e^{\mu(s-t) + \frac{\sigma^2 (s-t)}2} - \kappa\right)
\]
With $\mu=0$, $\sigma=0.5$, $\kappa=1.1$, $\tau=2$, and $t=1.5$, Figure \ref{fig:figure5} is the graph of $q_t(s)$ from $s=1.5=t$ to $s=2=\tau$. This option has the same parameters ($\mu, \sigma, \kappa, \tau$) as the one described in Figures \ref{fig:figure2} and \ref{fig:figure2++} above, except that it is entered into at time $t=1.5$ instead of time $0$. But whereas Figure \ref{fig:figure2++} indicates an estimated exercise date $\varsigma \approx 1.83$, Figure \ref{fig:figure5} indicates an exercise date $\varsigma_{1.5} =2$.

Thus the  estimated  value $w(t)$ (or $w_t$) of an American call option at time $t$ ($0\leq t <\tau$) is $w_t=$
\begin{equation}\label{wt}
  z_t \bN^{0,1}
\left( \frac{\ln \frac {z_t}{\kappa} + \left( \rho + \frac 12 \sigma^2\right) \left(\varsigma_t -t \right)}{\sigma \sqrt{\varsigma_t-t}} \right)
-\kappa e^{-\rho \left(\varsigma_t -t \right)} \bN^{0,1}
\left( \frac{\ln \frac {z_t}{\kappa} + \left( \rho - \frac 12 \sigma^2\right) \left(\varsigma_t -t \right)}{\sigma \sqrt{\varsigma_t-t}} \right)
\end{equation}
At time $t=\tau$ the value of an unexercised American call option is
\[
w_\tau\;\;=\;\; \max\{z_\tau - \kappa,\;0\}.
\]
The preceding investigation uses distribution function $\mathcal{G}^{\mu\sigma}$. But the assumption that price processes $Z$ follow a geometric Brownian distribution is generally false, as demonstrated in \cite{MTRV} Section 9.9, pages 479--485.

Also, both European and American options have exercise values which depend on the value taken by the underlying asset on the date of exercise of the option. In contrast, options of the \emph{Asian} type have value which depends, not just on the share value on the option exercise date, but on the average value of the share over the lifetime of the option:
\[
w(\tau) = \max\{A(z_T) - \kappa,\,\,0\},
\]
where $A(z_T)$
is the average of the known or observed values taken by the share in advance of the exercise date $\tau$, which, for this purpose, is the ``present''.

So if the lifetime of the option consists of one million ``time-ticks'', by time $\tau$ the share value $z(s)$ will have been observed for each of one million values of $z(s)$, and, with $T=\,]0,\tau]$, $A(z_T)$ is the sum of a million values of $z(s)$ divided by a million.

However, at any time $t$ earlier than $\tau$, including time $t=0$ when the option contract is created, the future values $z(s)$ ($0<s\leq \tau$) are unknown and unpredictable. But if their joint distribution $F$ is known, then $A(Z_T)$ is a joint-contingent observable,
\[
A(Z_T) \sq A(z_T)\left[\R_+^T, F\right],
\]
and likewise $f_\tau(Z_T) \sq \max\{A(Z_T) - \kappa,\,\,0\}\left[\R_+^T, F\right]$. The values $w_t$ of the Asian option at times $t$, $0\leq t <\tau$ can then be deduced from (8.47) of \cite{MTRV}, page 440:
\[
w_t = e^{-\rho(\tau -t)} \E^{\bar F}\left[f_\tau\left(\bar{Z}_T\right)\right]
= 
e^{-\rho(\tau -t)} \int_{\R_+^{]t, \tau]}}f_\tau(z_T)\bar{F}(I[N]).
\]
The defects of the hypotheses (assumption of geometric Brownian motion) underlying the preceding argument are outlined in Section 9.9 (pages 479--485) of \cite{MTRV}. Sections 9.10 and 9.11 of \cite{MTRV} outline an empirical method for estimating the actual joint distribution functions for such processes. 

The remainder of this article applies this method to estimate the value of an Asian call option on Glanbia shares, using the Maple program. 

 Both European and American options have exercise values which depend on the value taken by the underlying asset on the date of exercise of the option. In contrast, other kinds of option contract have value which depends, not just on the share value on the option exercise date, but on the values taken by the underlying asset at various times during the lifetime of the option contract.
 
For instance, an option of the \emph{Asian} type depends 
 on the average value of the share over the lifetime of the option:
\[
w(\tau) = \max\{A(z_T) - \kappa,\,\,0\},
\]
where $A(z_T)$
is the average of the known or observed values taken by the share in advance of the exercise date $\tau$.

So if the lifetime of the option consists of one million ``time-ticks'', when time $\tau$ is reached the share value $z(s)$ will have been observed for each of one million values of $z(s)$, and, with $T=\,]0,\tau]$, $A(z_T)$ is the sum of a million values of $z(s)$ divided by a million.

However, at any time $t$ earlier than $\tau$, including time $t=0$ when the option contract is created, the future values $z(s)$ ($0<s\leq \tau$) are unknown and unpredictable. But if their joint distribution $F$ is known, then $A(Z_T)$ is a joint-contingent observable,
\[
A(Z_T) \sq A(z_T)\left[\R_+^T, F\right],
\]
and likewise $f_\tau(Z_T) \sq \max\{A(Z_T) - \kappa,\,\,0\}\left[\R_+^T, F\right]$. The values $w_t$ of the Asian option at times $t$, $0\leq t <\tau$ can then be deduced from (8.47) of \cite{MTRV}, page 440:
\begin{equation}\label{value of Asian option 1}
w_t = e^{-\rho(\tau -t)} \E^{\bar F}\left[f_\tau\left(\bar{Z}_T\right)\right]
= 
e^{-\rho(\tau -t)} \int_{\R_+^{]t, \tau]}}f_\tau(z_T)\bar{F}(I[N]),
\end{equation}
in accordance with the risk-neutral martingale pricing arguments\footnote{Unlike the assumption of log-normality, these arguments seem to be fairly reliable.} of \cite{MTRV} Sections 8.13--8.19, pages 433--466, with 
\begin{equation}\label{terminal value of Asian option}
f_\tau(z_T) = \max\{A(z_T) - \kappa,\,\,0\}
\end{equation}
for $z_T \in \R_+^T$.
In the case of European and American options, $f_\tau$ depends on a single variable $x_\tau$ or $x_\varsigma$ with $\tau$ (or $\varsigma$) fixed, so the latter integral reduces to an integral on $\R_+^{\tau}$ (or $\R_+^{\varsigma}$). 

But the integral in
(\ref{value of Asian option 1})
latter integral is infinite-dimensional. Note that if $t>0$ is the ``present'', some values of $z_T, = \{z(s): 0 < s \leq \tau\}$, are known values, while the others are still in the future, and must be regarded as unpredictable observables. So
\[
w_t = 
e^{-\rho(\tau -t)} \int_{\R_+^{]t, \tau]}}f_\tau\left(
(z_{T'},z_{T''})
\right)\bar{F}(I[N]),
\]
where the values $z_{T'} \in \R_+^{]0,t]}$ are known and the values $z_{T''} \in \R_+^{]t,\tau]}$ are potential occurrences of the joint-basic observable
\[
Z_{T''} \sq z_{T''}\left[\R_+^{]t, \tau]},F\right].
\]
To sum up,
\begin{equation}
\label{Asian1}
w_0 = e^{-\rho(\tau )} \int_{\R_+^{]0, \tau]}}f_\tau(z_T)\bar{F}(I[N])
\;\;\;\mbox{ where}\;\;\;f_\tau(z_T) = \max\{A(z_T) - \kappa\}.
\end{equation}
Instead of assuming $F=\mathcal{G}^{\mu\sigma}$, pages 486--489 of \cite{MTRV} suggest an empirical approach using the risk-neutral, no-arbitrage pricing argument of \cite{MTRV} Section 8.14, pages 436--438. Applying this argument to the Glanbia share price series [224], the task can be tackled in steps such as the following.
\begin{itemize}
\item
Using the historic Glanbia data $z_T$, construct empirical distribution values $F(I[N])$ as in Section 9.11 of \cite{MTRV} pages 487--489.
\item
Estimate the growth rate $\mu$ of the price process. This can be done by, for example, calculating least squares regression for the Glanbia data $z_T$.
\item
Calculate a discounted version $z'(s) = e^{-\mu s} z(s)$ of the Glanbia price data.
\item
Using the growth rate estimate $\mu$, adjust the distribution values $F$ so that, with respect to amended values $\bar F$, the discounted prices $z'(s), w'(s)$ are occurrences of martingales $Z'_T\sq z'_T[\R_+^T, \bar F]$, $W'_T\sq w'_T[\R_+^T, \bar F]$:
\[
\E^{\bar F}[Z'_s] =z'(0),\;\;\;\;\;\;\E^{\bar F}[W'_s] =w'(0)  
\]
for all $s$, $0<s\leq \tau$; with $z'(0) = z(0)$, $w'(0) = w(0)$.
\item
Calculate an estimated value $w(0)$ by estimating the integral
\[
w(0) = w'(0) = \E^{\bar F}[W'_\tau] =
\int_{\R_+^{]0,\tau]}} w'(\tau) \bar{F}(I[N])
= \int_{\R_+^{]0,\tau]}} f_\tau (z'_T) \bar{F}(I[N]).
 \]
\end{itemize}
For Asian options the latter integral depends on all occurrences $z(s)$, $0<s \leq \tau$. So the integral is not cylindrical, and in order to estimate it, suitable  regular partitions\footnote{$(r,q)$-partitions are a bit cumbersome in practice, and other kinds of regular partitioning will be used below.} of $\R_+^{]0,\tau]}, = \R_+^T$, can be used, along with step function estimates of the integrand.

Step function approximations are, in principle at least, easy enough to understand and to apply---even in the case of multi-dimensional integrals, when we do \textbf{not} have recourse to the immense simplification that occurs when the integrand is a cylinder function whose integral happens to reduce to a one-dimensional integral, as in the cases of European and American options.

But, even when step function simplicity is present, it is still necessary to engage with the financial principles involved in risk-neutral valuation. This confronts us with plenty of tricky issues.

The main issue is converting the process distribution $F$ to  a martingale distribution $\bar F$. How can this be tackled in the case of the Glanbia price process, for instance?

Accepting that $F$ cannot be taken to be the geometric Brownian function $\mathcal G$, \cite{MTRV} Sections 9.10 and 9.11 (pages 486--489) suggests using a counting or ``relative frequency'' method to estimate $F$ for an observable $X_T \sq x_T[\R^T,F]$. 

This is problematic. Suppose an observable $X$ has distribution function $P$ on sample space $\R$, $X \sq x[\R,P]$, so a $P$-measurable set of potential occurrences  $A \subset \R$ has probability $P(A)$. The relative frequency idea implies that ``multiple occurrences'' of $A$, expressed as a fraction of total occurrences, provide an estimate of $P(A)$. Some justification of this estimate is provided by the Law of Large Numbers (\cite{MTRV} Section 5.9, pages 224--233).

The trouble with this is that ``multiple occurrences'' of a random variable $X$ (or of a process $Z_T$) are contrary to the usual mathematical understanding of a random variable, or of a process. A random variable/observable is generally understood as a mathematical description of a \textbf{single} future possible occurrence of something; that is, a single future observation or measurement which has multiple possible values, just one of which can actually ``occur''. 

So there are no ``multiple occurrences'' from which relative frequency can be calculated as estimate of probability. 

On the other hand, multiple occurrences, and relative frequency, can be somehow extracted from the Law of Large Numbers by postulating, not a single random variable with multiple occurrences, but multiple random variables, each with a single occurrence. 

But in order to connect this line of reasoning to the ``relative frequency approx\-imates to probability'' idea,  this family of random variables must consist of statistically independent random variables.

Glanbia (or any other) historic share price data constitute a single occurrence $z_T$ of a family of joint observables $\{Z_t: t \in T\}$, written as process $Z_T\sq z_T[\R_+^T, F]$. The distribution function $F(I[N])$ for the process can be thought of as consisting of transition probabilities for joint occurrences
\[
\left(z(t_j) \in I(t_j),\,1 \leq j \leq n\right), \,\,\,\,\,\,N=\{t_1, \ldots , t_n\}.
\]
Can relative frequency be used to estimate $F(I[N])$? This question is addressed in \cite{MTRV} pages 486--489. The idea is as follows. 

The Glanbia share price series begins with the end-of-day price 0.8 on Friday 8 March 1991, and continues with the daily price for each day of 5-day weeks until Wednesday 9 March 2011 when the end-of-day price was 3.692. Thus the series contains over 5,000 individual items. 

The price transitions in $I[N]$ of (\ref{value of Asian option 1})
are possible transitions in the future. Assume that, at any stage in the ``history'' (past, present, or future) of the data,  the likelihood of such a pattern of transitions is some more-or-less fixed amount for this particular series of share prices.

It is possible to count how many transitions in the historic or past data correspond to the set of possible future transitions represented by $I[N]$.
To convert this to a relative frequency,  divide this number by the total number of possible transitions corresponding to $I[N]$ in the past data. 
Provided the number of price transitions $n$ in the event $I[N]$ is small compared to 5,000, the preceding assumptions suggest\footnote{This can be criticized from various angles. One benefit, however, is that there is no presumption of independence. Any conditioning or dependence is picked up by the counting process.} that this relative frequency could be used as an approximation to $F(I[N])$. 

Next, to perform risk-neutral valuation, the (estimated) distribution function values $F(I[N])$ must be converted to (estimated) martingale distribution values $\bar{F}(I[N])$ for $I[N] \in \I(\R_+^T)$. How can this be done?

Consider a single share price $z_t$ at some fixed time $t$, $0<t\leq \tau$. For cells $I_t =\,]u,v] \in \I(\R_+)$, methods such as those described above can be used to find an estimated distribution function $F_t(I_t)$ for the observable $Z_t \sq z_t\left[\R_+, F_t\right]$,
\[
\pr\left(z_t \in I_t\right)=F_t(I_t),\;\;\;\;\;\;\;\;\; \mu_t =\E[Z_t] = \int_{\R_+} z_t F_t(I_t) = \int_0^\infty z_t d\pr.
\]
The basic observable $Z_t$ can be regarded as a contingent observable in $\R_+^T$, with $f(Z_T) = Z_t$, $f(Z_T) \sq f(z_T)[\R_+^T,F]$.  Then $f$ is a cylinder function, $f(z_T) = z_t$, $t$ fixed; and
\[
\E[Z_T] = \int_{\R_+^T}f(z_T) F(I[N])= \int_{\R_+} z_t F_t(I_t) =\mu_t
\]
since $F(I[N]) =F_t(I_t)$ for $N=\{t\}$, $I[N] = I_t$.

Given a price process $Z_T \sq z_T[\R_+^T, F]$, risk-neutral pricing theory requires the construction of a process $\bar{Z}_T \sq {z}_T[\R_+^T,\bar{F}]$ such that, under the distribution function $\bar F$, the discounted values $e^{-\rho s} z(s)$ are occurrences of a martingale, with 
\[
\E^{\bar F}\left[e^{-\rho s} \bar Z_s\right] = z(0),\;\;\;\;\mbox{ or }\;\;
\E^{\bar F}\left[ \bar Z_s\right] = e^{\rho s}z(0)
\]
for each $s \in T$, where $z(0)$ is the known initial price, and where, for present purposes, the risk-free interest rate $\rho$ is taken to be constant for all $s$.

\begin{example}
\label{bar F}
The construction of $\bar F_t$ can be illustrated as follows. Suppose we have an observable $X \sq x[\R,H]$, with mean value $\mu = \E[X]$. Suppose we wish, by changing the values of $H$, to construct a different observable $\bar{X}$ whose mean is some given number $\E^{\bar H}[\bar X] =\bar{\mu}, \neq \mu$. 
This can be done in two stages. First construct $X' \sq x'[\R,H']$ with mean $0$, then construct $\bar{X}=\bar{x}[\R, \bar{H}]$ with\footnote{The elements $x, x', \bar x$ are not different numbers; each of them represents an arbitrary possible occurrence in $\R$.} mean 
$\bar{\mu}$. 
For the first stage, suppose $I' =\,]u',v'] \in \R$. Write
\[
u=u'+\mu,\;\;v=v'+\mu,\;\;\;\mbox{ so }\;\;
u'=u-\mu,\;\;v'=v -\mu,
\]
and, for $I_t=\,]u,v]$, define
$
H'(I')=H(I)
$,
so $\E^{H'}[X'] =0$.
The second stage is similar. For $\bar I_t =\,]\bar u,\bar v] \in \R$ write
\[
u=\bar u +\mu -\bar{\mu},\;\;v=\bar v+\mu-\bar{\mu},\;\;\;\mbox{ so }\;\;
\bar u=u-\mu+\bar{\mu},\;\;\bar v=v -\mu +\bar{\mu};
\]
and, for $I_t=\,]u,v]$, define
$
\bar H(\bar I)=H(I)$, so $\E^{\bar H}[\bar X] = \bar \mu 
$, as required.
\end{example}
Returning to the construction of $\bar Z_T \sq \bar z_T[\R_+^T, \bar F]$, Example \ref{bar F} shows how to construct $\bar F_t$ for each $t \in T$, so, for $N = \{t\}$ the values $\bar F(I[N])$ can be found. 

But if $\bar Z_T$ is to be defined as a \emph{process} or \emph{joint} observable, the \emph{joint} distribution function $\bar F(I[N])$ must be defined, not just for singletons $N = \{t\}$, but for \emph{all} $N \in \N(T)$ and all corresponding $I[N] \in \I(\R_+^T)$. The counting procedure addresses by considering various patterns of transition.

The Glanbia daily price share data described in \cite{MTRV} Section 9.11 (pages 487--489) consists of 5225 daily end-of-day Glanbia share prices, beginning Friday 8 March 1991 and concluding Wednesday 9 March 2011.

Suppose the latter point in time is the present, and a 3 month call option contract on Glanbia shares is entered into at that moment. Since the share prices are reported daily on a 5-day week basis, take the 3-month term of the option to be 60 days, starting on morning of 10 March 2011 and expiring at end-of-day Wednesday 1 June 2011, a total of 5219 individual days/prices.

Suppose the option is the Asian type, with pay-off depending on the average share price over the term of the option.

To find an economic price for entering into this contract, it will be useful to have estimates of the daily expected value of a share during the 60-day term of the option; likewise the daily standard deviation of the share price. Also, joint probability values $F(I[N])$ need to be estimated for the share price process during the term of the option.

Maple calculations on the historic share prices can be used to make such estimates.

The historic Glanbia price data \cite{GlanbiaData} can be found in an Excel file in 

\noindent{\small{
\texttt{https://sites.google.com/site/stieltjescomplete/}
}}

The immediate aim is to produce estimates of expected share price value for each day of the term of the option---that is, the 
expected share price on 10 March 2011, on 11 March 2011, and so on, up to and including 1 June 2011; all of which are in the ``future''. 

Likewise, to estimate daily share price standard deviation for each day of the term of the option. Maple calculations on the historic or past data provides such estimates.

\begin{maple}\label{Maple code for empirical distribution value}
\end{maple}

\vspace{-7pt}

\begin{enumerate}
\item
{\small
\texttt{with(ExcelTools); 
with(Statistics);} 
\item
$Q := $
\texttt{Import("glanbia.xls")}:
\item
$A1 := $
\texttt{seq}$(Q[j][2], j = 2500 .. 5010):$
$ A2 := [A1]: A3 := $\texttt{Vector}$(A2):$
\item
 $J1 := $\texttt{seq}$(j, j = 2500 .. 5010); J2 := [J1]; J3 := $\texttt{Vector}$(J2)$; 
 \item
 $y := $\texttt{ExponentialFit}$(J3, A3, t)$; 
 \item
 \texttt{GlanbiaGraph} $:= $[\texttt{seq}$([t, Q[t+7][2]], t = 2500 .. 5000)]$; 
 \item
 \texttt{plot([GlanbiaGraph}, $y], t = 2500 .. 5000)$
\item
$p := 7; q := p+59$; 
\item
\texttt{for} $k$ \texttt{from 1 to 5160 do}
\item
$ A[k] :=$ \texttt{seq}$(Q[j][2], j = p .. q)$; 
\item
$B[k] := [A[k]]; p := p+1; q := q+1; $
\item
$m[k] :=$ \texttt{Mean}$(B[k]); s[k] :=$ \texttt{StandardDeviation}$(B[k])$: 
\item
\texttt{end do}; 
\item
$m :=$ \texttt{seq}$(m[k], k = 1 .. 5160); s := $\texttt{seq}$(s[k], k = 1 .. 5160)$; 
\item
$S :=$ \texttt{seq}$(s[k], k = 5000 .. 5160); T := [S]; s0 :=$ \texttt{Mean}$(T)$
\item
$M[1] := 3.7*$\texttt{exp}$(0.006); M[2] := 3.7*$\texttt{exp}$(0.012); M[3] := 3.7*$\texttt{exp}$(0.018)$:
\item
\texttt{for }$k$ \texttt{from} $1$ \texttt{to} $5219$ \texttt{do}
\item
$ z[k] := Q[k+6][2]:$ 
\item
\texttt{end do:}
\item
$N := 6; J := seq(j, j = -3 .. 2)$; \texttt{seq}$(q, i = 1 .. 3)$; \texttt{seq}$(b, i = 1 .. 3)$; 
\item
\texttt{for} $i$ \texttt{from 1 to 216 do}
\item
 \texttt{for} $j$ \texttt{from 1 to 3 do}
\item 
 $ q[j] :=$ \texttt{trunc}$((i-1)/N^(j-1))+1$; 
 \item
 $b[j] :=$ \texttt{mod}$(q[j]-1, N)+1$; 
\item 
 $p[i][j] := J[b[j]]$ 
 \item
 \texttt{end do}
\item 
 \texttt{ end do}
\item
$ex := 3.5; w0 := 0;$ 
\item
\texttt{Prob} $:= 0$; 
\item
\texttt{for} $i$ \texttt{from 1 to 216 do }
\item
$f := 0$; 
\item
\texttt{for} $k$ \texttt{ from 4550 to 5000 do }
\item
\texttt{if}
$(z[k+20] >= m[k+20]+s[k+20]*p[i][1]$
\item
\texttt{and}
$ z[k+20] < m[k+20]+(p[i][1]+1)*s[k+20])$
\item
\texttt{and} 
$z[k+40] >= m[k+40]+s[k+40]*p[i][2])$
\item
\texttt{and} 
$z[k+40] < m[k+40]+(p[i][2]+1)*s[k+40])$
\item
\texttt{and}
$ z[k+60] >= m[k+60] +p[i][3]*s[l+60]$
\item
\texttt{and} $z[l+60] < m[l+60]+(p[i][3]+1)*s[l+60])$
\item
\texttt{ then} $f := f+1$: 
\item
\texttt{end if};
\item
\texttt{ end do}; 
\item
$P := (1/50)*f$; 
\item
\texttt{Prob} $:=$ \texttt{Prob}$+P$; 
\item
$a1 := (M[1]+p[i][1]*s0 + M[2] +p[i][2]*s0 +M[3]+p[i][3]*s0)/3-ex$; 
\item
\texttt{if} $a1 > 0$ \texttt{then} $a2 := a1$; \texttt{end if}; 
\item
\texttt{if} $a1 <= 0$ \texttt{then} $a2 := 0$; \texttt{end if}; 
\item
$a3 := a2*$\texttt{exp}$(-0.02)$; 
\item
$w := a3*P$; 
\item
$w0 := w0+w$: 
\item
\texttt{end do}; 
\item
$w0$; 
\item
\texttt{evalf(Prob)};}
\end{enumerate}
For the purpose of explanation, the lines of Maple code above are numbered. These numbers are not part of the program and should not be included in actual Maple code.

Lines 1--7 take the historic Glanbia data and import them into the program for the purpose of performing Maple calculations on them. These data can be found in an Excel file in
\noindent{\small{
\texttt{https://sites.google.com/site/stieltjescomplete/}
}} 

\noindent and should be stored in the same folder in which the above Maple code is located, in order for this code to be able to access them. In version 18 of Maple, the file with \texttt{Import} command must, on first use, be closed and then re-opened before \texttt{Import} will work successfully.

Lines 3 and 4 convert the share price data to a format suitable for line 5, using about half of the available prices---from day 2500 to day 5010; it is a matter of judgement which data to include.

Line 5 calculates 
a least squares regression for the price data, in the form $y=a \exp( \mu t)$. The idea here is that the prices follow some underling growth rate $\mu$ of a ``proportional'' character; with random variation of prices superimposed on this underlying trend.  In other words, if the share price on day $t$ is $z_t$, and disregarding the superimposed random variation of prices, then 
\[
\frac{z_{t+1}}{z_t} \mbox{ is constant}, \;\;\;\;
\ln \left(\frac{z_{t+1}}{z_t}\right) = \mu.
\]
In this case the exponential best fit calculated by Maple is  approximately
\[
y=0.1 e^{0.0007t},
\]
so daily growth rate is approximately $\mu = 0.0007$. This translates into an annual growth rate of the share values amounting to about 20\%.

\begin{figure}
\begin{minipage}[h]{0.45\linewidth}
\centering
\includegraphics[width=\textwidth]{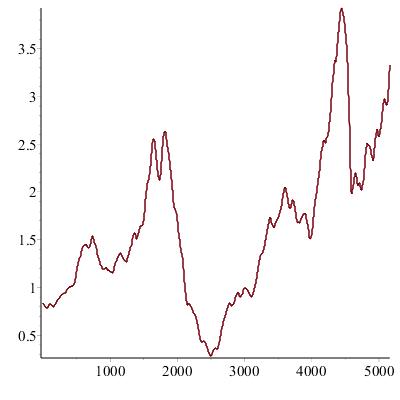}
\caption{All moving averages}
\label{fig:All moving averages}
\hspace{0.5cm}
\centering
\includegraphics[width=\textwidth]{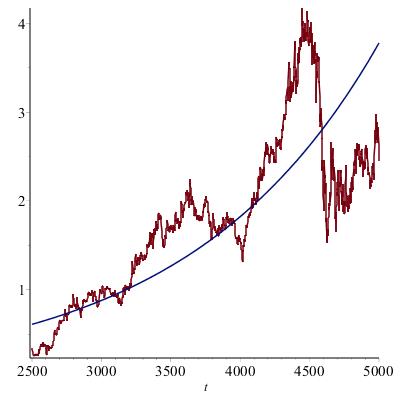}
\caption{Exponential regression}
\label{regression}

\end{minipage}
\hspace{0.5cm}
\begin{minipage}[h]{0.45\linewidth}
\centering
\includegraphics[width=\textwidth]{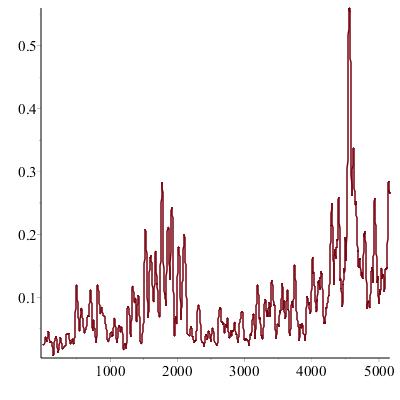}
\caption{All moving SD's}
\label{fig:All moving SD's}
\hspace{0.5cm}
\centering
\includegraphics[width=\textwidth]{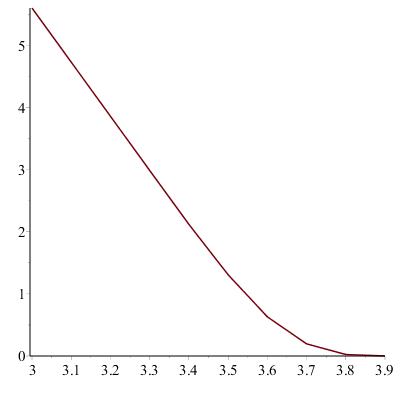}
\caption{Option values}
\label{Asian option values}
\end{minipage}
\end{figure}

Lines 7 and 8 of the Maple code produce a graph ($y$) of this underling growth, superimposed on a graph of the actual prices. See Figure \ref{regression} at the end of this article.

For the purpose of pricing an Asian option entered into ''today'' (9 March 2011), when ``today's'' share price is 3.7, a risk-free daily growth rate of 0.0003 (which is less than the underlying growth rate $\mu=0.0007$), so the underlying trend\footnote{Not the actual trend, just the hypothetical risk-free trend needed for martingale pricing of discounted values. Line 16 of the code shows these successive ``risk-free trend'' values at intervals of 20 days during the term of the option.} values of the share during the period of the Asian option (up to 1 June 2011, or 60 days in total) is taken to be 
\[
y=0.1 e^{0.0003t}.
\]
Then, provided we can superimpose on these price trend values the appropriate amount of random variability, we can estimate the value of the Asian option by using (\ref{value of Asian option 1}). 

Lines 8--13 are the first step in accomplishing this. This part of the program calculates successive 60-day average values of the historic share prices. Also the corresponding standard deviations. The latter demonstrate the scale of the random variability or volatility in these share prices, and are a kind of indicator of the joint likelihood distribution $F(I[N]$ that ``determines'' the share price process $Z_T \sq z_t[\R_+^T,F]$.

These ``moving averages'' and ``moving standard deviations'' are illustrated in Figures \ref{fig:All moving averages} and \ref{fig:All moving SD's} at the end of this article.

The Maple program will also use these mean-and-standard-deviation data as ``$3\sigma$'' partition points in a regular partition of $\R_+^T$,  $T$ being the 60-day period or term of the Asian option, in order to estimate the option valuation (\ref{value of Asian option 1}). 

Line 15 takes an average $s0$ of the 60-day standard deviations. This is to be taken as the standard deviation of the daily share price for each day of the 60-day term of the Asian option---see line 44.

Line 16 computes three 20-day trend values $M[1], M[2], M[3]$ for the 60-day term of the option. But these are not ``true'' trend values; they are the hypothetical values showing a risk-free daily growth rate of 0.0003 needed to carry out the risk-neutral martingale valuation of (\ref{value of Asian option 1}).

Lines 17 to 19 ascribe the familiar notation $z[k]$ (or $z_t$) to the historic share price values. Lines 20--27 compute the permutations, with repetition, of the six numbers $-3,-2,-1,0,1,2$ taken three at a time. There are $6^3 =216$ such permutations, which, for $i=1$ to 216, are each given by
\[
(p[i][1], \;\;\;\;p[i][2],\;\;\;\;p[i][3]).
\]
One such permutation is $(1,-1,2)$; that is
\[
p[i][1]=1,\;\;p[i][2]=-1,\;\;p[i][3]=2.
\]
With mean $m$ and standard deviation $s$, these numbers are then used to construct $3\sigma$ partitioning intervals
\[
[m-p[i][j]*s, m-(p[i][j]+1)*s[,
\]
which, for $p[i][j]=-3,-2, \ldots , 2$, give six cells

\[
[m-3s,m-2s[,\;\;\;[m-2s,m-s[,\;\;\;[m-s,m[,\newline
[m,m+s[,\;\;\;[m+s,m+2s[,\;\;\;[m+2s,m+3s[.
\]

On any particular day the share price will usually be found to lie in one of these intervals. The probability of $|z_t -m| > 3s$ is small, and can be ignored in the option valuation calculation. Take the start of 9 March 2011 to be time $t=0$, and 1 June 2011 to be $t=\tau$, and $T=\,]0, \tau]$, with $t=\tau_1 =20$, $t=\tau_2 =40$, and $t=\tau=60$. 
For $j=1,2,3$, write
\begin{equation} \label{partitioning cells 1}
I_j = [M[j]+p[i][j]*s0,\;\;\;M[j]+(p[i][j]+1)*s0[
\end{equation}
where the values $s0$ and $M[j]$ are given by lines 15 and 16 of the Maple code.

Then, with
$N=\{\tau_1, \tau_2, \tau\}$ the domain $\R_+^T$ of the option is partitioned by cylinders 
\begin{equation}  \label{partitioning cells 2}
I[N] = \prod_{j=1}^3I_j \times \R_+^{T \setminus M}.
\end{equation}
There are 216 such cylindrical cells, corresponding to the 216 permutations $(p[i][1],p[i][2],p[i][3])$. This is a regular partition in $\R_+^T$. It does not fully exhaust the domain $\R_+^T$, but the probability that $z_T$ is not in one of these cylindrical intervals is very small.
Line 28 of the Maple program sets the exercise price of the option at $ex=3.5$. The assignment $w0:=0$ sets the initial value of the Riemann sum estimate of (\ref{value of Asian option 1}). The Riemann sum has 216 terms, one for each permutation $(p[i][j]; j=1,2,3)$. Each term of the Riemann sum has value $w$ (line 48), and the term values $w$ are accumulated as $w0$ in the 216 iterations of lines 48 and 49.

The ultimate value returned by $w0$ is the value of the Riemann sum estimate of (\ref{value of Asian option 1}). For option exercise price $ex=3.5$, the Maple program gives initial value $w0=1.31$. This is the $w_0$ or $w(0)$ of (\ref{value of Asian option 1}). The expression \texttt{Prob} of line 29 is used in the code to keep track of, or accumulate, the probabilities generated in the calculation. Total probability returned by line 43 is approximately 0.97, indicating that the $3\sigma$ regular partition of $\R_+^T$, though non-exhaustive, is almost full.

Using the historic Glanbia share price data, lines 30--43 of the program use a ``relative frequency''-type argument to produce estimates of the joint transition probabilities $F(I[N]$ needed to calculate the Riemann sum estimates $w0$ of  (\ref{value of Asian option 1}). 

Line 32 shows that this  program only uses 50 cycles or iterations of the $3 \times 20$-day  transitions in the partitioning cells (\ref{partitioning cells 2}). For greater accuracy, it is easy to increase this sample size. 

The relative frequency calculation for a single permutation of transitions $(p[i][j]: j=1,2,3)$ is done in line 42. This value of $P$ is the estimate of $F(I[N]$ for a single term of the Riemann sum for (\ref{value of Asian option 1}).

For $i=1, \ldots , 216$, each permutation $(p[i][j]: j=1,2,3)$ generates a single term of the Riemann sum for (\ref{value of Asian option 1}). The calculation of each such term is done in lines 44--48.

How are the values $F(I[N])$ converted to risk-neutral probabilities $\bar F(I[N])$? 
This is accomplished in line 44. The quantities $M[j] + p[i][j] *s0$ are the lower bounds of partitioning cells of the form (\ref{partitioning cells 1}) and (\ref{partitioning cells 2}), and these are taken as tag points or evaluation points for the function $A(z_T)$ in (\ref{value of Asian option 1}). 

But these boundary points are unlike the corresponding boundaries in lines 33--38. The latter are growing at the daily trend growth rate $\mu=0.0007$, whereas the bounds in line 44 are
growing at the risk-free growth rate $\rho = 0.0003$. 

So, in some sense, the probabilities $F(I[N])$ produced from lines 33-42 are, in line 44 applied to the ``wrong'' cells of $\R_+^T$. Except that this is the appropriate adjustment needed to produce the hypothetical $\bar F(I[N])$ in $\R_+^T$.

The point-integrand in (\ref{value of Asian option 1}) is the function (\ref{terminal value of Asian option}), and this is calculated as $a2$ in lines 44--46. The discounted\footnote{For 60-day discounting, the daily rate 0.0003 becomes approximately $60 \times 0.0003 = 0.018$. Line 47 uses $e^{-0.02}$ instead of $e^{-0.018}$ for discounting at the risk-free rate.} value of (\ref{terminal value of Asian option}) is calculated  in line 47. Multiplying this by $P=\bar F(I[N])$ gives the value of a single term $w$ of the Riemann sum. Summing over all the permutations, $i=1, \ldots , 216$, gives the Riemann sum estimate of the option value at initial time $t=0$:
$
w0=w_0=w(0) = 1.3.$

To test out the above Maple code, it is advisable to test its component parts individually in order to more easily detect and correct transcription and coding errors. Also, at the relevant points in the code, alternative parameters and estimation tactics can be experimented with.

In the above code the average $A(z_T)$ was estimated at only three 20-day intervals. For a 60-day option the Glanbia share price data in the Excel file permit up to 60 daily price values to be used to calculate $A(z_T)$. In practice, a reasonable balance must be struck between the demands of accuracy and the scale of computing power available. The latter quickly escalates when more accuracy is demanded.


\noindent
\cite{MTRV}, page 480, has a twenty-year graph of the Glanbia share prices. Figure \ref{regression} is a graph of part of the data, with superimposed exponential regression graph $y=\exp(0.0007 t)$.

The first average calculated by Lines 8--13 of the Maple program is the mean of the prices for day 1 (8 March 1991) up to and including day 60. The second average is for day 2 (9 March 1991) up to and including day 61 of the data. The final average is the mean of the prices on day 5160 of the data
up to and including day 5219 (9 March 2011). 

This is a total of 5160 cycles, giving 5160 means and 5160 standard deviations. The latter are displayed in Figure \ref{fig:All moving SD's}.

Since the first moving average is the average for the first 60 days, it can be applied to day 30 or day 31, half-way through the first cycle. Likewise for each of the other moving averages; and also the standard deviations.

The graph of moving averages, Figure \ref{fig:All moving averages}, is a ``smoothed out'' version of the graph (Figure \ref{regression}) of the original Glanbia share price data. But it is not as smooth as the exponential regression graph for the data, in Figure \ref{regression}.

Figure \ref{Asian option values} shows how the initial value of the Asian option depends on the terminal exercise price. A low exercise price of 3.0 results in high option price of about 5.5. A high exercise price of 3.8 produces a low option price of about 0.02.

\end{document}